\documentclass[a4paper,12pt]{article}
\pdfoutput=1
\usepackage{graphicx}
\usepackage{mathtools}
\usepackage{amssymb}
\usepackage{amsfonts}
\usepackage[footnotesize]{caption}
\usepackage[font=scriptsize]{subcaption}
\usepackage{color}
\usepackage{braket}
\usepackage{cite}
\usepackage{hyperref}
\usepackage{url}
\usepackage{multirow}
\usepackage{relsize}
\usepackage{fullpage}
\usepackage{makecell}
\usepackage{blkarray}
\usepackage{fullpage}

\usepackage[utf8]{inputenc}

\newcommand{\newc}{\newcommand}
\newc{\be}{\begin{equation}}
\newc{\ee}{\end{equation}}
\newc{\bea}{\begin{eqnarray}}
\newc{\eea}{\end{eqnarray}}
\newc{\simlt}{~\mbox{\smaller\(\lesssim\)}~}
\newc{\simgt}{~\mbox{\smaller\(\gtrsim\)}~}

\newcommand{\pmatr}[1]{\begin{pmatrix} #1 \end{pmatrix}}

\begin{document}

\begin{titlepage}
\begin{center}
{\bf\Large
\boldmath{
$R_{K^{(*)}}$ with leptoquarks and the origin of Yukawa couplings
}
} \\[12mm]
Ivo~de~Medeiros~Varzielas$^{\dagger}$
\footnote{E-mail: \texttt{ivo.de@udo.edu}},
Stephen~F.~King$^{\star}$%
\footnote{E-mail: \texttt{king@soton.ac.uk}},
\\[-2mm]
\end{center}
\vspace*{0.50cm}
\centerline{$^{\dagger}$ \it
CFTP, Departamento de F\'isica, Instituto Superior T\'ecnico, Universidade de Lisboa,}
\centerline{\it Avenida Rovisco Pais 1, 1049 Lisboa, Portugal}
\centerline{$^{\star}$ \it
School of Physics and Astronomy, University of Southampton,}
\centerline{\it
SO17 1BJ Southampton, United Kingdom }

\vspace*{1.20cm}

\begin{abstract}
{\noindent
We construct a model where the Yukawa couplings of the Standard Model (SM) fermions arise from the breaking of a $Z_2$ symmetry and through mixing with a fourth family of vector-like fermions. By adding a scalar leptoquark,
which is an electroweak triplet and odd under the $Z_2$ symmetry, we obtain an explanation for $R_{K^{(*)}}$ that is linked to the origin of the Yukawa couplings. The coupling of SM fermions to the leptoquark is mediated by the fourth family fermions, and is predicted to be related to CKM entries and mass ratios of SM fermions.
}
\end{abstract}
\end{titlepage}

\section{Introduction}

There has been 
mounting evidence for semi-leptonic $B$ decays violating $\mu - e$ universality in excess of the rates predicted by the Standard Model (SM)
\cite{Descotes-Genon:2013wba,Altmannshofer:2013foa,Ghosh:2014awa}.
The LHCb Collaboration and other experiments have observed anomalies in $B\rightarrow K^{(*)}l^+l^-$
decays such as the $R_K$ \cite{Aaij:2014ora} and $R_{K^*}$ \cite{Aaij:2017vbb} ratios of $\mu^+ \mu^-$ to $e^+ e^-$ final states, 
which are observed to be about $70\%$ of their expected values with a $4\sigma$ deviation from the SM,
and the $P'_5$ angular variable,
not to mention the $B\rightarrow \phi \mu^+ \mu^-$ mass distribution in $m_{\mu^+ \mu^-}$.

Following the measurement of $R_{K^*}$ \cite{Aaij:2017vbb}, a number of phenomenological analyses of these data, 
see e.g. \cite{Glashow:2014iga, Calibbi:2015kma, Descotes-Genon:2015uva, Capdevila:2017bsm, DAmico:2017mtc, Hiller:2017bzc, Geng:2017svp, Ciuchini:2017mik, Ghosh:2017ber, Alok:2017jaf, Alok:2017sui, Bardhan:2017xcc},
favour a new physics operator of the form 
$\bar b_L\gamma^{\mu} s_L \, \bar \mu_L \gamma_{\mu} \mu_L$,
or of the form, $\bar b_L\gamma^{\mu} s_L \, \bar \mu \gamma_{\mu} \mu$,
each with a coefficient $\Lambda^{-2}$ where $\Lambda \sim 31.5$ TeV,
or some linear combination of these two operators. 
Interestingly, the first of these operators may result from an electroweak triplet scalar leptoquark $S_3$ with couplings \cite{Hiller:2017bzc},
\be
\lambda^{ij}S_3Q_iL_j\equiv \lambda^{ij}S_3^{\beta \gamma}Q^{\alpha}_i(i\sigma_2)^{\alpha \beta} L_j^{\gamma}
\label{S3}
\ee
where the $SU(2)_L$ indices $\alpha, \beta, \gamma $ are suppressed
on the left-hand side, with $Q_i$ and $L_j$ with $i,j=1,2,3$ denoting the three families of quark and lepton electroweak doublets
in a two component Weyl notation. The operator 
$\bar b_L\gamma^{\mu} s_L \, \bar \mu_L \gamma_{\mu} \mu_L$ 
(along with other operators) is generated via tree-level $S_3$ exchange after a Fierz transformation.

We have proposed 
an explanation of $R_{K^{(*)}}$ based on a fourth 
vector-like family which carries gauged $U(1)'$ charge, where the anomalies cancel between conjugate representations in the fourth family \cite{King:2017anf}, and the $U(1)'$ is spontaneously broken to yield a massive $Z'$ gauge boson at the TeV scale. 
In the ``fermiophobic'' example, we showed that, even if the three chiral families do not carry $U(1)'$ charges, 
effective non-universal couplings, as required to explain $R_{K^{(*)}}$, can emerge indirectly due to 
mixing with the fourth family \cite{King:2017anf}. 
We have pursued the original idea \cite{King:2017anf} in several subsequent papers, including:
F-theory models with non-universal gauginos \cite{Romao:2017qnu};
$SO(10)$ models (including the question of neutrino mass) \cite{Antusch:2017tud};
$SU(5)$ models (focussing on the Yukawa relation $Y_e\neq Y_d^T$)  \cite{CarcamoHernandez:2018aon};
and flavourful $Z'$ portal models with a coupling to a fourth-family singlet Dirac neutrino Dark Matter,
including an extensive discussion of all phenomenological constraints
\cite{Falkowski:2018dsl}. A closely related idea was also considered some time later \cite{Raby:2017igl},
including other phenomenological implications such as 
the muon $g-2$ and $\tau \rightarrow \mu \gamma$.

Despite the huge recent literature on $R_{K^{(*)}}$,
relatively few papers are concerned with its possible connection with Yukawa couplings \cite{Varzielas:2015iva, Pas:2015hca, Hiller:2017bzc, Guo:2017gxp, Aloni:2017ixa,King:2018fcg, Hati:2018fzc}.
For example, recently we investigated 
the possible connection between the experimental signal for new physics in 
$R_{K^{(*)}}$ and the origin of fermion Yukawa couplings in a $Z'$ model \cite{King:2018fcg}.
We focussed on $Z'$ models where the physics responsible for the $Z'$ mass and couplings is also responsible for 
generating the effective Yukawa couplings, 
providing a link between flavour changing observables and 
the origin of Yukawa couplings.

The starting point of \cite{King:2018fcg} was the model proposed in \cite{Ferretti:2006df, Calibbi:2008yj},
which involves one vector-like family distinguished by a discrete $Z_2$.
We replaced the $Z_2$
by a new $U(1)'$ gauge group, spontaneously broken at the TeV scale, under which the three chiral families 
are neutral but the vector-like fourth family is charged.
The mixing between the fourth family and the three chiral families 
provided the Yukawa couplings,
as well as the non-universal effective $Z'$ couplings involving the three light families, 
providing a link between the Yukawa and $Z'$ couplings.
However any such $Z'$ model is subject to a strong constraint from $\Delta m_{B_s}$, which is generated by tree-level
$Z'$ exchange. This compares to leptoquark models which only contribute to $\Delta m_{B_s}$ via a one loop diagram,
which is therefore naturally suppressed.

In the present paper our starting point is again the model proposed in \cite{Ferretti:2006df, Calibbi:2008yj},
which involves one vector-like family distinguished by a discrete $Z_2$, where 
the Yukawa couplings of the Standard Model (SM) fermions arise from the breaking of the $Z_2$ and through mixing with the fourth family of vector-like fermions. Here we introduce a scalar leptoquark
$S_3$, which is an electroweak triplet and odd under the $Z_2$ parity, and obtain an explanation for $R_{K^{(*)}}$ that is linked to the origin of the Yukawa couplings. The couplings of SM fermions to the leptoquark in Eq.~\ref{S3} in our model
have a particular structure, mediated by the fourth family vector-like fermions,
with the couplings predicted to be related to Cabibbo-Kobayashi-Maskawa (CKM) entries and mass ratios of SM fermions.

The layout of the remainder of the paper is as follows.
In Section \ref{sec:model} we present the model and discuss the Yukawa couplings in a convenient basis.
In Section \ref{sec:basis} we derive the leptoquark couplings in the mass basis.
We analyse the associated leptoquark phenomenology of the model 
in Section \ref{sec:pheno}. We conclude in Section \ref{sec:conc}.

\section{The model \label{sec:model}}

The model is analogous to the setup in \cite{King:2018fcg}, and is defined by the field content and respective symmetry assignments listed in Table~\ref{tab:funfields1}. In addition to the SM symmetries, we add a $Z_2$ symmetry.
Discrete symmetries such as this can arise from various different UV completions, e.g. by orbifolding in extra dimensions.

The scalar content includes one $SU(2)_L$ doublet $H$, a SM singlet scalar $\phi$, and a scalar leptoquark $S_3$ that is an anti-triplet of $SU(3)_c$ and a triplet of $SU(2)_L$. Beyond the SM fermions, we add 3 families of $Z_2$ neutral right-handed (RH) neutrinos (complete singlets), and a fourth family of vector-like fermions for each type of fermion. 

\begin{figure}[ht]
\centering
	\includegraphics[scale=0.2]{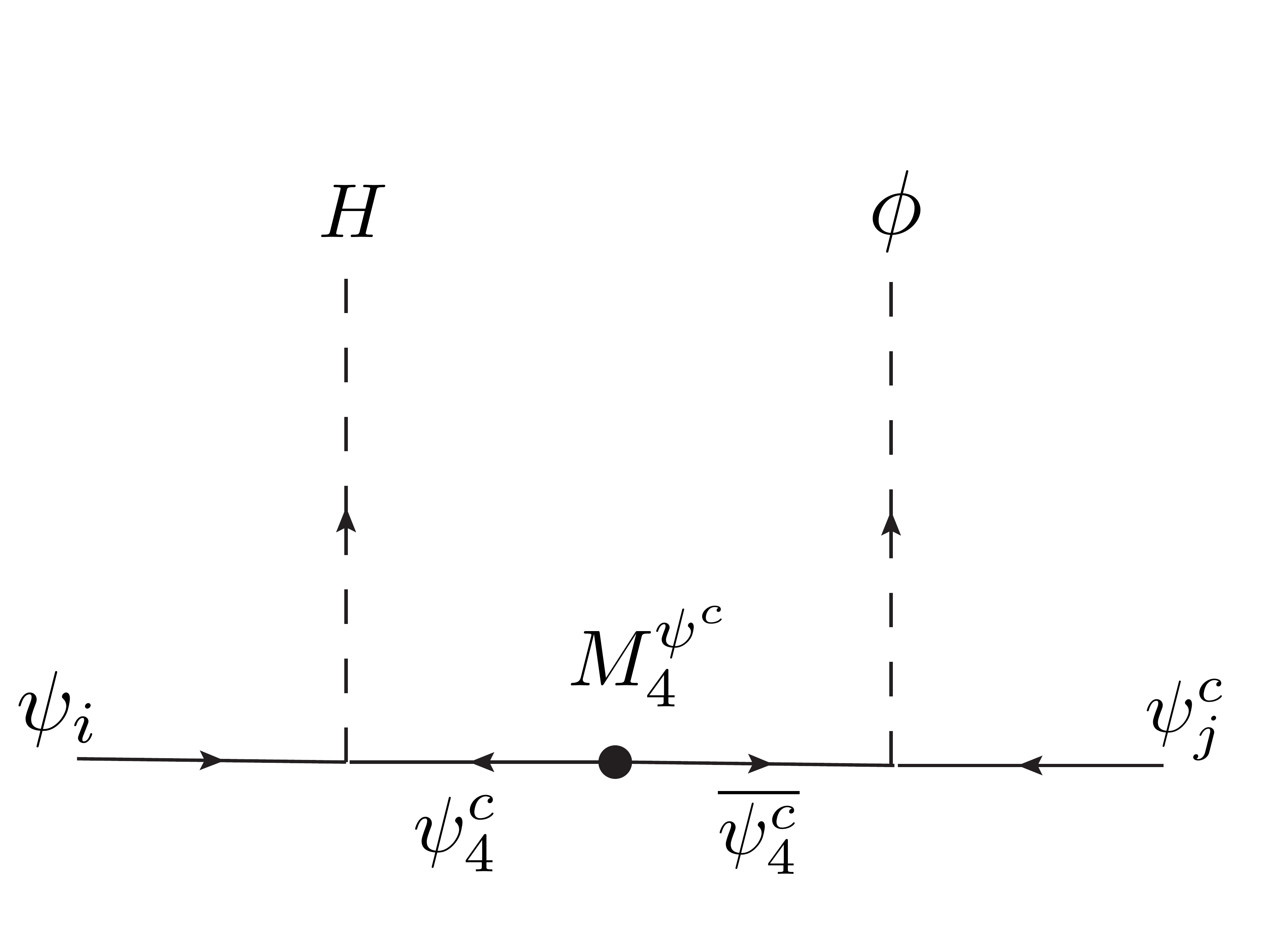}
\hspace*{1ex}
	\includegraphics[scale=0.2]{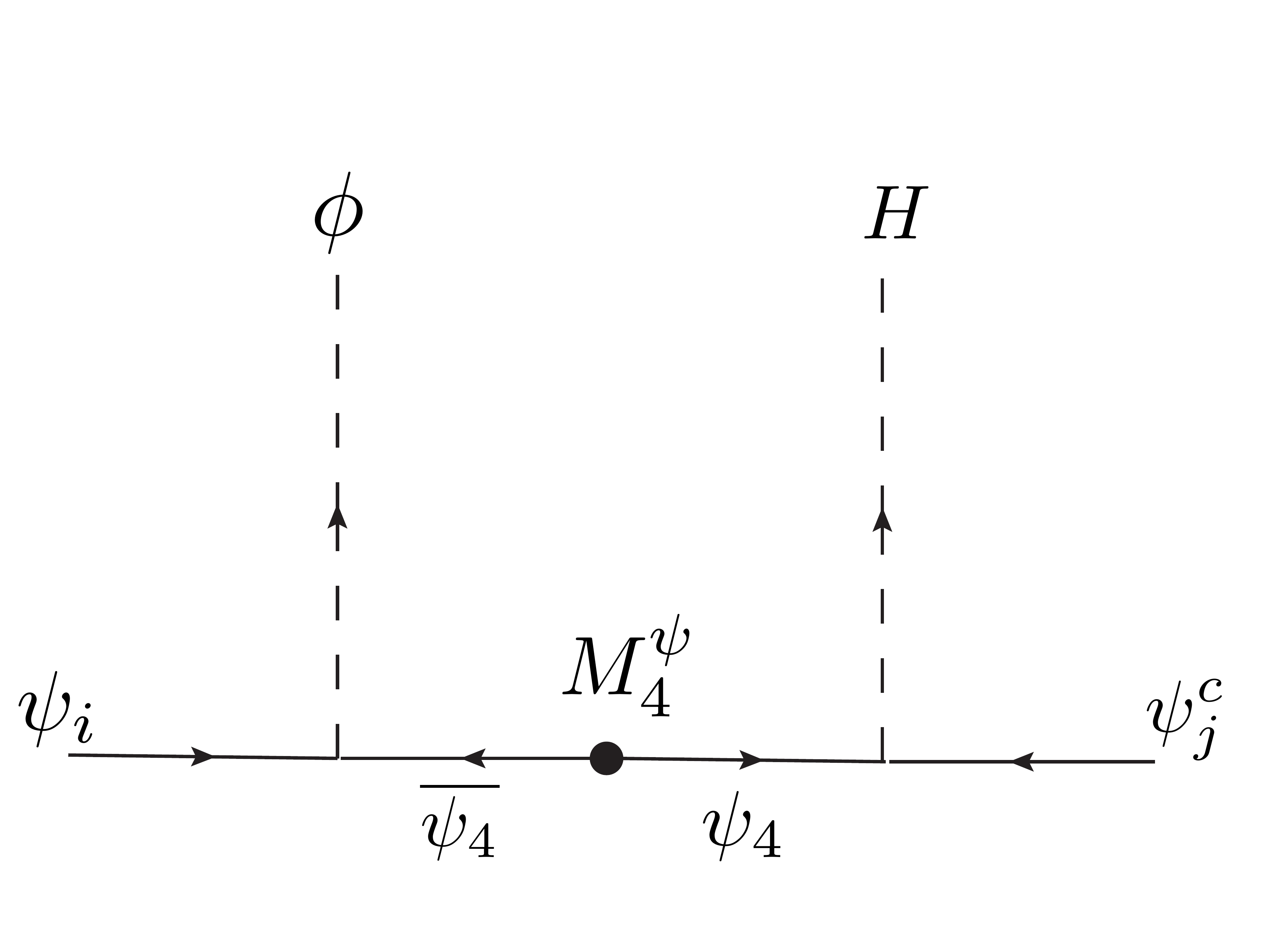}
\caption{Diagrams in the model which lead to the effective Yukawa couplings in the mass insertion approximation.}
\label{Fig1}
\end{figure}

Although SM fermions are neutral under the $Z_2$ symmetry, renormalisable Yukawa couplings are only enabled through the couplings $\phi$ and to the fourth family fermions, because $H$ is odd under $Z_2$. The effective Yukawa couplings arise from diagrams shown in Fig.~\ref{Fig1}. 

The leptoquark $S_3$ has renormalisable couplings to fourth family fermions
\be
\lambda^{L_4}_{Q_i} S_3  Q_i L_4+ \lambda^{Q_4}_{L_i} S_3 L_i Q_4 \,,
\label{lambda}
\ee
(in left-handed Weyl notation). Notice that the coupling $S^{\dagger}_3 Q_4 Q_4$ is forbidden by $Z_2$ \footnote{Leptoquark models can face issues with fast proton decay, see e.g. \cite{Assad:2017iib} and references therein.}.
The effective leptoquark couplings to SM fermions arise from Fig.~\ref{Fig2}, via the renormalisable leptoquark couplings. In compact notation we write:
\bea
{\cal L}^{ren} &=&
y^{\psi}_{i4}H  \psi_i {\psi^c_4} 
+  y^{\psi}_{4i}H {\psi_4} \psi^c_i+x^{\psi}_{i}\phi \psi_i \overline{\psi_4} 
+ x^{\psi^c}_{i}\phi \psi^c_i \overline{\psi^c_4}\nonumber \\
&+& M^{\psi}_{4}\psi_4 \overline{\psi_4}
+ M^{\psi^c}_{4}\psi^c_4 \overline{\psi^c_4}
+\lambda^{L_4}_{Q_i} S_3  Q_i L_4+ \lambda^{Q_4}_{L_i} S_3 L_i Q_4 \,,
\label{Lag_ren}
\eea
where $\psi_i$ ($i=1,2,3$) here denotes the left-handed fields $Q_i$ and $L_i$ and $\psi_i^c$ denotes the 
(CP conjugated) right-handed fields $u^c_{i}$, $d^c_{i}$ and $e^c_{i}$ that appear explicitly in Table \ref{tab:funfields1}, while those with the subscript $4$,
$\psi_4$, $\psi^c_4$, $\overline{\psi_4}$, $\overline{\psi^c_4}$, 
denote the vector-like fourth family fields which appear as internal ``messenger" fields in Fig.~\ref{Fig1}.

The $Z_2$ symmetry is broken by the vacuum expectation value (VEV)
$\langle \phi \rangle$, assumed to be around the scale of the masses $M^{\psi}_{4}$, $M^{\psi^c}_{4}$,
whose values are put in by hand.

\begin{table}
\centering
\begin{tabular}{| l c c c c |}
\hline
Field & $SU(3)_c$ & $SU(2)_L$ & $U(1)_Y$ &$Z_2$\\ 
\hline \hline
$Q_{i}$ 		 & ${\bf 3}$ & ${\bf 2}$ & $1/6$ & $+1$ \\
$u^c_{i}$ 		 & ${\overline{\bf 3}}$ & ${\bf 1}$ & $-2/3$ & $+1$\\
$d^c_{i}$ 		 & ${\overline{\bf 3}}$ & ${\bf 1}$ & $1/3$ & $+1$\\
$L_{i}$ 		 & ${\bf 1}$ & ${\bf 2}$ & $-1/2$ & $+1$\\
$e^c_{i}$ 		 & ${\bf 1}$ & ${\bf 1}$ & $1$ & $+1$\\
$\nu^c_{i}$         & ${\bf 1}$ & ${\bf 1}$ & $0$ & $+1$\\
\hline
\hline
$Q_{4}$ 		 & ${\bf 3}$ & ${\bf 2}$ & $1/6$ & $-1$\\
$u^c_{4}$ 		 & ${\overline{\bf 3}}$ & ${\bf 1}$ & $-2/3$ & $-1$\\
$d^c_{4}$ 		 & ${\overline{\bf 3}}$ & ${\bf 1}$ & $1/3$ & $-1$\\
$L_{4}$ 		 & ${\bf 1}$ & ${\bf 2}$ & $-1/2$ & $-1$\\
$e^c_{4}$ 		 & ${\bf 1}$ & ${\bf 1}$ & $1$ & $-1$\\
$\nu^c_{4}$         & ${\bf 1}$ & ${\bf 1}$ & $0$ & $-1$\\
\hline
\hline
$\overline{Q_{4}}$ 		 & $\overline{{\bf 3}}$ & $\overline{{\bf 2}}$ & $-1/6$ & $-1$\\
$\overline{u^c_{4}}$ 		 & ${{\bf 3}}$ & ${\bf 1}$ & $2/3$ & $-1$\\
$\overline{d^c_{4}}$ 		 & ${{\bf 3}}$ & ${\bf 1}$ & $-1/3$ & $-1$\\
$\overline{L_{4}}$ 		 & ${\bf 1}$ & $\overline{{\bf 2}}$ & $1/2$ & $-1$\\
$\overline{e^c_{4}}$ 		 & ${\bf 1}$ & ${\bf 1}$ & $-1$ & $-1$\\
$\overline{\nu^c_{4}}$         & ${\bf 1}$ & ${\bf 1}$ & $0$ & $-1$\\
\hline
\hline
$\phi$ & ${\bf 1}$ & ${\bf 1}$ & $0$ &$-1$ \\
\hline
\hline
$S_3$ &  $\overline{\bf 3}$ & ${\bf 3}$ & $1/3$ & $-1$\\
\hline
\hline
$H$ & ${\bf 1}$ & ${\bf 2}$ & $-1/2$ &$-1$ \\
\hline
\end{tabular}
\caption{The model consists of 
three left-handed chiral families $\psi_i=Q_i,L_i$ and 
$\psi^c_i =u^c_i,d^c_i,e^c_i,\nu^c_i$ 
($i=1,2,3$),
plus a fourth vector-like family consisting of $\psi_4, \psi^c_4$ plus 
$\overline{\psi_4},\overline{\psi^c_4}$, together with the $Z_2$ breaking scalar fields
$\phi$ and the Higgs scalar doublet $H$ which is odd under $Z_2$ (together with its CP conjugate $\tilde{H}$).
It also involves a scalar leptoquark $S_3$.}
\label{tab:funfields1}
\end{table}

In the quark sector we are allowed to 
go to a particular basis in $Q_i, d^c_i, u^c_i$ ($i=1,\ldots 3$) flavour space where 
$x^{Q}_{1,2}=0$, $y^{u}_{41,42}=0$, $y^{d}_{41,42}=0$ (in the notation of \cite{King:2018fcg}).
Then we can further rotate the first and second families to set $x^{u^c}_1=0$, $x^{d^c}_1=0$ and $y^{u}_{14}=0$
(but in general $y^{d}_{14}\neq 0$ since the quark doublet rotations have already been used up).
In this basis the matrix of quark couplings are
\be
	 \pmatr{
	&u^c_1& u^c_2&u^c_3&u^c_4&\overline{Q_4}\\ 
	\hline
	Q_1|&0&0&0&0 &0\\
	Q_2|&0&0&0&y^{u}_{24}\tilde{H} &0\\
	Q_3|&0&0&0&y^{u}_{34}\tilde{H} &x^{Q}_{3}\phi\\ 
	Q_4|& 0 & 0 & y^{u}_{43}\tilde{H}&0&M^{Q}_{4}\\ 
	\overline{u^c_4}|&0&x^{u^c}_{2}\phi&x^{u^c}_{3}\phi& M^{u^c}_{4}&0},
	 \pmatr{
	&d^c_1& d^c_2&d^c_3&d^c_4&\overline{Q_4}\\ 
	\hline
	Q_1|&0&0&0&y^{d}_{14}H &0\\
	Q_2|&0&0&0&y^{d}_{24}H &0\\
	Q_3|&0&0&0&y^{d}_{34}H &x^{Q}_{3}\phi\\ 
	Q_4|& 0 & 0 & y^{d}_{43}H&0&M^{Q}_{4}\\ 
	\overline{d^c_4}|&0&x^{d^c}_{2}\phi&x^{d^c}_{3}\phi& M^{d^c}_{4}&0} \,.
	\label{M^quark_an_basis}
\ee

We consider analogously the charged leptons, which are (like the quarks) hierarchical. In the respective special basis in $L_i, e^c_i$ ($i=1,\ldots 3$) flavour space where 
$x^{L}_{1,2}=0$, $y^{e}_{41,42}=0$, as well as $x^{e^c}_1=0$, and $y^{e}_{14}=0$:
\be
	 \pmatr{
	&e^c_1& e^c_2&e^c_3&e^c_4&\overline{L_4}\\ 
	\hline
	L_1|&0&0&0&0 &0\\
	L_2|&0&0&0&y^{e}_{24}H &0\\
	L_3|&0&0&0&y^{e}_{34}H &x^{L}_{3}\phi\\ 
	L_4|& 0 & 0 & y^{e}_{43}H&0&M^{L}_{4}\\ 
	\overline{e^c_4}|&0&x^{e^c}_{2}\phi&x^{e^c}_{3}\phi& M^{e^c}_{4}&0} \,.
	\label{M^lepton_an_basis}
\ee

To understand the origin of Yukawa couplings in this model, consider the two diagrams in Fig.~\ref{Fig1}. They show that the effective Yukawa couplings are the sum of two contributions - one that depends on the left-handed messengers $M^{\psi}_4$
(with mass $M^{Q}_4$ for the quarks or $M^{L}_4$ for the leptons), and one that depends on the (CP conjugated) right-handed messengers $M^{\psi^c}_4$. 
As indicated by Eqs.\ref{M^quark_an_basis}, \ref{M^lepton_an_basis}, the effective Yukawa matrices from left-handed and from right-handed messengers will both have a first column of zeros, corresponding to zero couplings to $\psi_1^c$,
so we conclude the first family will have zero mass without further modification.
Dropping one of the two diagrams in Fig.~\ref{Fig1},
leads to the 
second family becoming massless as well. 
This is suggestive of a natural explanation of the smallness of the second family masses compared to the third family masses, namely that 
one term dominates over the other one. 
This was named ``messenger dominance'' in  \cite{Ferretti:2006df, Calibbi:2008yj}. 
To account for small $V_{cb}$ in the quark sector, it is natural to assume that 
the left-handed quark messengers dominate over the right-handed messengers, $M_4^Q\ll M_4^{d^c},M_4^{u^c}$,
which was named ``left-handed messenger dominance'' in  \cite{Ferretti:2006df, Calibbi:2008yj},
with the further assumption $M_4^Q\ll M_4^{d^c}\ll M_4^{u^c}$ reproducing the stronger mass hierarchy
in the up sector (compared to the down sector). Assuming all this leads to $|V_{cb}|\sim m_s/m_b$ 
with $V_{ub}$, though naturally small, being unconstrained \cite{Ferretti:2006df, Calibbi:2008yj}.
However to explain the smallness of the Cabibbo angle requires further model building 
such as an $SU(2)_R$ symmetry \cite{Ferretti:2006df, Calibbi:2008yj}, although here we assume its smallness is accidental.

To accurately determine the Yukawa matrices, 
we now follow \cite{King:2018fcg} and perform the diagonalisation of the matrices in Eqs.\ref{M^quark_an_basis}, \ref{M^lepton_an_basis},
in the approximation that $\langle H \rangle \ll \langle \phi \rangle$ (but without requiring $\langle \phi \rangle \ll M_4^{Q,L}$)
 in order to find the three massless eigenstates which we identify with the SM fermions. 
 To be precise, what we mean by ``diagonalisation'' is that the matrices are rotated to a form in which 
 the first three rows and columns of Eqs.\ref{M^quark_an_basis}, \ref{M^lepton_an_basis} contain zeros (in the limit that $\langle H\rangle$ is neglected).
 This diagonalisation procedure is more rigorous than the mass insertion approximation, leading to structures compatible with the above qualitative conclusions that were derived from the diagrams in Fig.~\ref{Fig1}.
The ``diagonalisation'' of Eqs.\ref{M^quark_an_basis}, \ref{M^lepton_an_basis} requires the mixing angles,
\bea
\sin \theta_{34}^Q &= s^{Q}_{34} &= \frac{x^Q_3 \langle \phi \rangle }{ \sqrt{(x^Q_3 \langle \phi \rangle )^2+(M^{Q}_{4})^2 }} \,, \label{rigorous_angle}\\
\sin \theta_{34}^L &= s^{L}_{34} &= \frac{x^L_3 \langle \phi \rangle }{ \sqrt{(x^L_3 \langle \phi \rangle )^2+(M^{L}_{4})^2 }} \,,
\eea
and respective cosines $c^Q_{34}$, $c^L_{34}$, as well as
\bea
\theta^{u^c}_{24} &\approx \frac{x_2^{u^c}\langle \phi \rangle }{M_4^{u^c}} \,, \\
\theta^{d^c}_{24} &\approx \frac{x_2^{d^c}\langle \phi \rangle }{M_4^{d^c}} \,, \\
\theta^{e^c}_{24} &\approx \frac{x_2^{e^c}\langle \phi \rangle }{M_4^{e^c}} \,,
\eea
and
\bea
\theta^{u^c}_{34} &\approx \frac{x_3^{u^c}\langle \phi \rangle }{M_4^{u^c}} \,, \\
\theta^{d^c}_{34} &\approx \frac{x_3^{d^c}\langle \phi \rangle }{M_4^{d^c}} \,, \\
\theta^{e^c}_{34} &\approx \frac{x_3^{e^c}\langle \phi \rangle }{M_4^{e^c}} \,.
\eea
The 
effective Yukawa coupling matrices are then read off from the upper $3\times3$ blocks of Eqs.\ref{M^quark_an_basis}, \ref{M^lepton_an_basis} 
after the above ``diagonalisation'':
\be
	{y}^u_{ij}= \pmatr{
		0&0&0 \\
	0& \theta^{u^c}_{24}y^{u}_{24}& \theta^{u^c}_{34} y^{u}_{24}\\
	0& c^{Q}_{34} \theta^{u^c}_{24}y^{u}_{34}& c^{Q}_{34} \theta^{u^c}_{34}y^{u}_{34}  }
	+\pmatr{
		0&0&0 \\
	0&0&0 \\
	0&0&s^{Q}_{34} y^{u}_{43} },
\ee
\be
		 {y}^d_{ij}=  \pmatr{
		0& \theta^{d^c}_{24}y^{d}_{14}& \theta^{d^c}_{34} y^{d}_{14}\\
	0& \theta^{d^c}_{24}y^{d}_{24}& \theta^{d^c}_{34} y^{d}_{24}\\
	0& c^{Q}_{34} \theta^{d^c}_{24}y^{d}_{34}& c^{Q}_{34} \theta^{d^c}_{34}y^{d}_{34}  }
		 +
	\pmatr{
		0&0&0 \\
	0&0&0 \\
	0&0&s^{Q}_{34} y^{d}_{43}   }	 \,,
\ee
\be
	{y}^e_{ij}= \pmatr{
		0&0&0 \\
	0& \theta^{e^c}_{24}y^{e}_{24}& \theta^{e^c}_{34} y^{e}_{24}\\
	0& c^{L}_{34} \theta^{e^c}_{24}y^{e}_{34}& c^{L}_{34} \theta^{e^c}_{34}y^{e}_{34}  }
	+\pmatr{
		0&0&0 \\
	0&0&0 \\
	0&0&s^{L}_{34} y^{e}_{43} } \,.
\ee
Note that in each case there are two contributions which depend on the right-handed messenger masses and left-handed messenger masses, respectively,
as we discussed above.

We now rewrite these $3\times3$ Yukawa in a compact form
\be
y^{u,e}_{ij}=  \pmatr{0 & 0 & 0 \\
0 & y^{u,e}_{22} & y^{u,e}_{23}\\
0 & y^{u,e}_{32} & y^{u,e}_{33}
} \,,
\label{Yuk_effective_ue}
\ee
\be
y^{d}_{ij}=  \pmatr{0 & y^{d}_{12} & y^{d}_{13} \\
0 & y^{d}_{22} & y^{d}_{23}\\
0 & y^{d}_{32} & y^{d}_{33}
} \,,
\label{Yuk_effective_d}
\ee
where we stress that the $y_{ij}$ couplings where $i,j=1,2,3$ (but not $4$) are effective Yukawa couplings obtained as functions of the renormalisable $x$ couplings, the $y_{i4}$, $y_{4j}$ couplings and ratios of $\langle \phi \rangle$ to the respective messenger masses $M_4$. 

Due to left-handed messenger dominance, $y^{u,d,e}_{33}$ dominate over the other entries (due to the contributions with $M_4^{Q,L}$), leading to hierarchical masses for the third family of SM fermions and allowing the use of small angle approximation in diagonalising the Yukawa. We now introduce $\theta_{ij}$ with $i,j=1,2,3$ (but not $4$) which are the angles involved in the diagonalisation of the Yukawa coupling matrices with the effective couplings $y_{ij}$, in Eqs.\ref{Yuk_effective_ue}, \ref{Yuk_effective_d}
\be
\theta_{23}^{u,d,e} \simeq y^{u,d,e}_{23}/y^{u,d,e}_{33} \,,
\ee
and further for the down quarks
\be
\theta_{13}^{d} \simeq y^{d}_{13}/y^{d}_{33} \,,
\ee
\be
\theta_{12}^{d} \simeq y^{d}_{12}/y^{d}_{22} \,.
\ee
At this level, the remaining mixing angles are zero (similarly to the masses of the first family). Given that the hierarchy of the charm to top mass (governed by $y^{u}_{22}/y^{u}_{33} \sim M^Q_4/M^{u^c}_4$) is stronger than the hierarchy of the strange to bottom mass (governed by $y^{d}_{22}/y^{d}_{33} \sim  M^Q_4/M^{d^c}_4$), we conclude also that $y^{d}_{23}/y^{d}_{33} \gg y^{u}_{23}/y^{u}_{33}$. The contribution to the CKM mixing angles from the up sector is negligible, and we can take to a reasonable approximation that, in the special basis we are working so far, the up quark Yukawa couplings are already (approximately) diagonal. With this reasonable assumption, we are basically considering that the family eigenstate, $Q_3$, contains the top (the mass eigenstate, $t$) and the down-type counterpart decomposed with CKM elements. The same applies for $Q_{2,1}$, which contain respectively the charm $c$ and the up $u$ and their down-type counterparts.

In the lepton sector, in the absence of the complete model for the neutrinos, we simply assume that the charged leptons Yukawa couplings are diagonalized by a small $\theta_{23}^e \sim m_\mu/m_\tau$.
We note that the phenomenology discussed later is independent of the details of the model of neutrino mass.

For the neutrinos, due to $\nu_i^c$ being neutral under SM and $Z_2$, large Majorana mass terms are allowed and lead to the seesaw mechanism. As the charged lepton Yukawa couplings are approximately diagonal in this basis, the Pontecorvo-Maki-Nakagawa-Sakata (PMNS) matrix comes mostly from neutrino contributions that arise after the seesaw. Nevertheless, the small rotation in the charged lepton sector controls the admixture of $\mu$ and $\tau$ contained in $L_3$.

To summarise, this model does not account for the PMNS angles (due to unspecified contribution from the neutrino sector, where the seesaw mechanism takes place) nor for the size of the Cabibbo angle (which is ultimately related to the ratio $y^d_{14}/y^d_{24}$). The model does account for the smallness of the remaining CKM angles, which are suppressed by the ratio of $M_4^Q/M_4^{d^c}$.

\section{Leptoquark couplings  \label{sec:basis}}

\begin{figure}[ht]
\centering
	\includegraphics[scale=0.2]{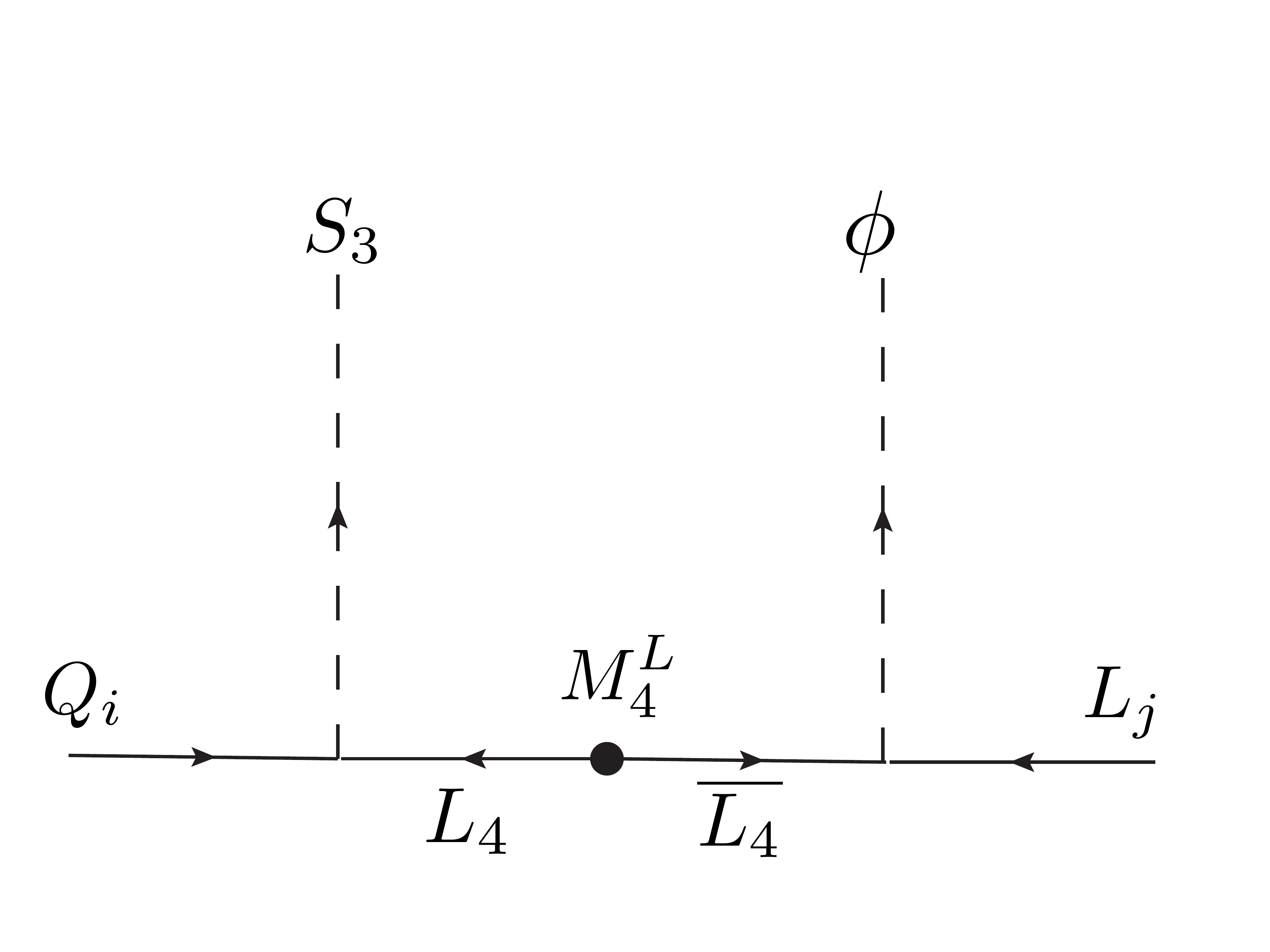}
\hspace*{1ex}
	\includegraphics[scale=0.2]{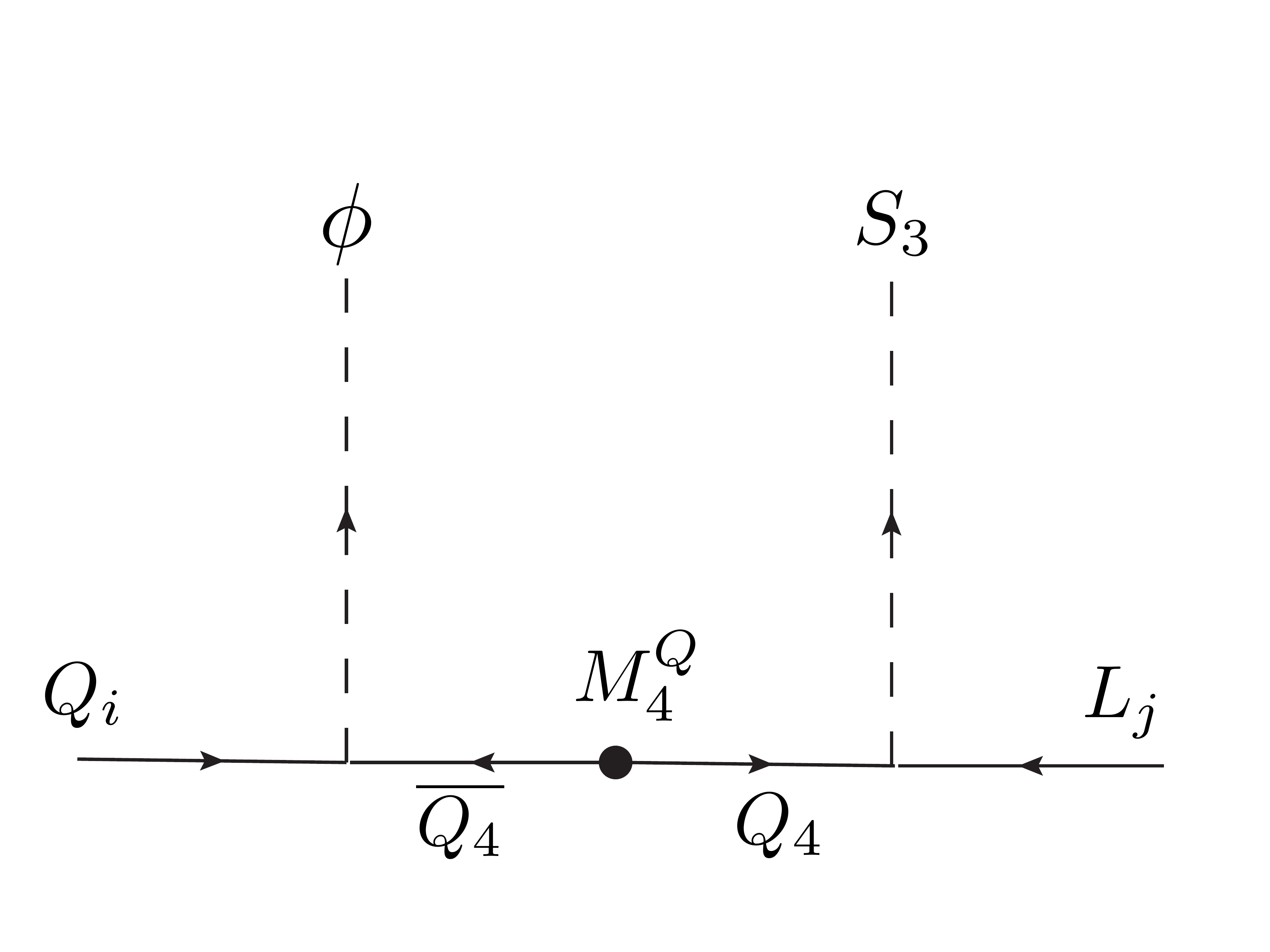}
\caption{Diagram in the model which leads to the effective $S_3$ couplings in the mass insertion approximation.}
\label{Fig2}
\end{figure}

In this section we discuss the effective leptoquark couplings arising from the diagrams in Fig.~\ref{Fig2}.
For such leptoquark couplings, we use the mass insertion approximation in the basis of Eqs.\ref{M^quark_an_basis}, \ref{M^lepton_an_basis},
without ``diagonalising'' these matrices.
However, we will eventually need to diagonalise the Yukawa matrices in Eqs.\ref{Yuk_effective_ue}, \ref{Yuk_effective_d},
and hence express the leptoquark couplings in the mass basis of the quarks and leptons.

In the basis of Eqs.\ref{M^quark_an_basis}, \ref{M^lepton_an_basis},
the effective leptoquark couplings from Fig.~\ref{Fig2} are
\be
\frac{x^{L}_{3}\langle \phi \rangle }{M^{L}_{4}} \lambda^{L_4}_{Q_i} S_3Q_iL_3 + 
 \frac{x^{Q}_{3}\langle \phi \rangle }{M^{Q}_{4}}
\lambda^{Q_4}_{L_i} S_3Q_3 L_i  \  \equiv  \lambda^{ij}S_3Q_iL_j \,,
\label{leptoquark_couplings}
\ee
where 
\bea
\label{leptoquark_family_basis}
 \lambda^{ij} = & \frac{x^{L}_{3}\langle \phi \rangle }{M^{L}_{4}}
\left[
\lambda^{L_4}_{Q_1} 
\left(
\begin{array}{ccc}
0 & 0 & 1 \\
0 & 0 & 0 \\
0 & 0 & 0
\end{array}
\right)
+
\lambda^{L_4}_{Q_2} \left(
\begin{array}{ccc}
0 & 0 & 0 \\
0 & 0 & 1 \\
0 & 0 & 0
\end{array}
\right)
+
\lambda^{L_4}_{Q_3} \left(
\begin{array}{ccc}
0 & 0 & 0 \\
0 & 0 & 0 \\
0 & 0 & 1
\end{array}
\right)
\right] \nonumber
\\
+&
 \frac{x^{Q}_{3}\langle \phi \rangle }{M^{Q}_{4}} \left[
\lambda^{Q_4}_{L_1} \left(
\begin{array}{ccc}
0 & 0 & 0 \\
0 & 0 & 0 \\
1 & 0 & 0
\end{array}
\right)
+
\lambda^{Q_4}_{L_2}
\left(
\begin{array}{ccc}
0 & 0 & 0 \\
0 & 0 & 0 \\
0 & 1 & 0
\end{array}
\right)
+
\lambda^{Q_4}_{L_3}
\left(
\begin{array}{ccc}
0 & 0 & 0 \\
0 & 0 & 0 \\
0 & 0 & 1
\end{array}
\right)
\right] \,.
\eea
The effective couplings with $\frac{x^{Q}_{3}\langle \phi \rangle }{M^{Q}_{4}} \lambda^{Q_4}_{L_i}$ are expected to be order unity, since the top quark Yukawa is large and demands
a large mixing with the $\phi$ fields.
Comparatively, the effective couplings with $\frac{x^{L}_{3}\langle \phi \rangle }{M^{L}_{4}} \lambda^{L_4}_{Q_i}$ are expected to be suppressed.

Notice that the leptoquark couples to specific combination of quarks and to specific combination of leptons, which in this basis where Eqs.\ref{M^quark_an_basis}, \ref{M^lepton_an_basis} applies, consists in just the family eigenstate $Q_3$ for the $M_4^Q$ suppressed term and just the family eigenstate $L_3$ for the $M_4^L$ suppressed term. 

The leptoquark couplings in Eq.\ref{leptoquark_family_basis}, involving $Q_i$ and $L_i$,
are in the basis where the Yukawa matrices are given
in Eqs.\ref{Yuk_effective_ue}, \ref{Yuk_effective_d}. 
In order to determine the leptoquark couplings to the quark and lepton mass eigenstates, we need to 
diagonalise the  Yukawa matrices in Eqs.\ref{Yuk_effective_ue}, \ref{Yuk_effective_d}, then express the leptoquark couplings
in the mass eigenstate basis. However, as discussed previously, the present model predicts the up-type quark mixing angles
to be small compared to the down-type mixing angles, so as an approximation, in the following we shall assume that the up-type Yukawa matrix to be diagonal,
and hence the left-handed down-type mixing angles are just the CKM mixing angles.

We now express $Q_i$ and $L_i$ (appearing in Eq.\ref{leptoquark_couplings}) in terms of the mass eigenstates, 
in the leading order approximation where we considered up-type quark Yukawa couplings to be diagonal.
In this approximation, 
$Q_3=(u_3,d_3)$ contains $u_3 = t_L$ which coincides with the top quark (mass eigenstate) and the down-type combination within the $SU(2)_L$ doublet is obtained by the CKM matrix, namely $d_3 = V_{td} d_L +  V_{ts} s_L + V_{tb} b_L$. Similar considerations for $Q_i=(u_i,d_i)$
lead us to the expressions
\bea
\label{dmass}
u_1 &\approx  u_L\,, \quad d_1 =& V_{ud} d_L +  V_{us} s_L + V_{ub} b_L \,, \nonumber \\
u_2 &\approx  c_L\,, \quad d_2 =& V_{cd} d_L +  V_{cs} s_L + V_{cb} b_L \,, \nonumber \\
u_3 &\approx  t_L\,, \quad d_3 =& V_{td} d_L +  V_{ts} s_L + V_{tb} b_L \,.
\eea
where the fields on the right-hand sides are in the mass eigenstate basis, while those of the left-hand side are in the basis $Q_i=(u_i,d_i)$
of Eq.\ref{leptoquark_couplings}.

In the charged lepton sector, the hierarchy is not as strong as in the up quarks, and we consider that $L_3=(\nu_3, e_3)$ contains $e_3$ which is 
an admixture of the $\tau_L$ and $\mu_L$ (mass eigenstates). According to Eq.\ref{Yuk_effective_ue}, $L_3$ does not contain any 
component of the left-handed electron $e_L$, namely $e_3 \simeq \theta_{23}^e \mu_L + (1 - (\theta_{23}^e)^2) \tau_L$.
In order to be consistent, we take $L_{2}$ as the orthogonal combination, thus leading us to
\bea
\label{emass}
e_1&=& e_L \,, \nonumber \\
e_2&\simeq& (1 - (\theta_{23}^e)^2) \mu_L - \theta_{23}^e \tau_L \,, \nonumber \\
e_3&\simeq& \theta_{23}^e \mu_L + (1 - (\theta_{23}^e)^2) \tau_L \,.
\eea

We are interested primarily in the Yukawa couplings of the leptoquark to down-type quark and charged lepton parts. In order to rewrite the effective leptoquark couplings $\lambda^{ij}S_3Q_iL_i$ in Eq.~\ref{leptoquark_couplings}
we first consider the $SU(2)_L$ components of $S_3$:
\be
\label{leptoquark_SU2}
S_3 =
\left(
\begin{array}{cc}
S_3^{1/3} & \sqrt{2} S_3^{4/3} \\
\sqrt{2} S_3^{-2/3} & -S_3^{1/3}
\end{array}
\right) \,,
\ee
in the notation of \cite{Hiller:2018wbv} where the superscripts are the electric charges of the components.
We may then expand the leptoquark couplings to the different species in Eq.~\ref{leptoquark_couplings},
\be
\lambda^{ij}S_3Q_iL_j=
\lambda^{ij} \left(\sqrt{2} d_i e_j S^{4/3}_{3} + d_i \nu_j S^{1/3}_{3} + \sqrt{2} u_i \nu_j S^{-2/3}_{3} - u_i e_j  S^{1/3}_{3}  \right) \,.
\label{lambda_dl}
\ee
We focus now on the $d_i$, $e_j$ couplings
 and bring the couplings from the basis of Eq.~\ref{leptoquark_family_basis}, into the mass basis of the charged leptons and down quarks in Eqs.~\ref{dmass}, \ref{emass},
 \be
\label{leptoquark_massbasis}
( \sqrt{2}S^{4/3}_{3} ) \lambda^{ij} d_i e_j =
( \sqrt{2}S^{4/3}_{3} )
\left(
\begin{array}{ccc}
d_L & s_L & b_L  \\
\end{array}
\right)
\lambda_{dl}
\left(
\begin{array}{c}
e_L   \\
\mu_L    \\
\tau_L 
\end{array}
\right) \,,
\ee
where we have defined the matrix of physical leptoquark couplings:
\be
\label{leptoquark_massbasis_1}
\lambda_{dl} \equiv
\left(
\begin{array}{ccc}
\lambda_{de} & \lambda_{d\mu} & \lambda_{d\tau}  \\
\lambda_{se}  & \lambda_{s\mu}  & \lambda_{s\tau}     \\
\lambda_{be}  & \lambda_{b\mu}   & \lambda_{b\tau}   
\end{array}
\right) \,.
\ee

To obtain our model predictions for this matrix of physical leptoquark couplings, we need to convert the couplings from Eq.\ref{leptoquark_couplings}.
We have in total 6 unknown couplings, 3 $\lambda^{L_4}_{Q_i}$ and 3 $\lambda^{Q_4}_{L_i}$. We note that the former contributions are suppressed by the larger $M_4^L \gg M_4^Q$, the relative suppression is $\sim m_\tau/m_t$.

The different $\lambda_{de}$, ..., $\lambda_{b\tau}$ in Eq.\ref{leptoquark_massbasis_1} are obtained from decomposing the $Q_i$ and $L_i$ flavour eigenstates in terms of the mass eigenstates according to Eqs.~\ref{dmass},\ref{emass}. Dropping the terms $(\theta_{23}^e)^2 \sim (m_\mu/m_\tau)^2$
we find:
\bea
\label{leptoquark_model}
\lambda_{dl} = \frac{x^{L}_{3}\langle \phi \rangle }{M^{L}_{4}}
\left[
\lambda^{L_4}_{Q_1} 
\left(
\begin{array}{ccc}
0  &  \theta_{23}^e V_{ud} & V_{ud} \\
0  &  \theta_{23}^e V_{us} & V_{us} \\
0  &  \theta_{23}^e V_{ub} & V_{ub}
\end{array}
\right)
+
\lambda^{L_4}_{Q_2} \left(
\begin{array}{ccc}
0  &  \theta_{23}^e V_{cd} & V_{cd} \\
0  &  \theta_{23}^e V_{cs} & V_{cs} \\
0  &  \theta_{23}^e V_{cb} & V_{cb}
\end{array}
\right)
+
\lambda^{L_4}_{Q_3} \left(
\begin{array}{ccc}
0  &  \theta_{23}^e V_{td} & V_{td} \\
0  &  \theta_{23}^e V_{ts} & V_{ts} \\
0  &  \theta_{23}^e V_{tb} & V_{tb}
\end{array}
\right)
\right] \nonumber
\\
+
 \frac{x^{Q}_{3}\langle \phi \rangle }{M^{Q}_{4}} \left[
\lambda^{Q_4}_{L_1} \left(
\begin{array}{ccc}
V_{td} & 0 & 0 \\
V_{ts} & 0 & 0 \\
V_{tb} & 0 & 0
\end{array}
\right)
+
\lambda^{Q_4}_{L_2}
\left(
\begin{array}{ccc}
0  & V_{td} & -\theta_{23}^e V_{td} \\
0  & V_{ts} & -\theta_{23}^e V_{ts} \\
0  & V_{tb} & -\theta_{23}^e V_{tb}
\end{array}
\right)
+
\lambda^{Q_4}_{L_3}
\left(
\begin{array}{ccc}
0  &  \theta_{23}^e V_{td} & V_{td} \\
0  &  \theta_{23}^e V_{ts} & V_{ts} \\
0  &  \theta_{23}^e V_{tb} & V_{tb}
\end{array}
\right)
\right] \,,
\eea
which can be compared to Eq.\ref{leptoquark_family_basis}.
We note again that the couplings involving $M_4^L$ are expected to be smaller by $m_\tau/m_t$. As an example, in Section \ref{sec:pheno}, we safely neglect the contribution
$\frac{x^{L}_{3}\langle \phi \rangle }{M^{L}_{4}} \lambda^{L_4}_{Q_2} \theta_{23}^e V_{cs}$ to $\lambda_{s\mu}$, with respect to the dominant contribution to this entry, which is $\frac{x^{Q}_{3}\langle \phi \rangle }{M^{Q}_{4}} \lambda^{Q_4}_{L_2} V_{ts}$.

\section{Phenomenology \label{sec:pheno}}

We now recast different phenomenological constraints that apply in general to models with the leptoquark $S_3$, in terms of our model predictions, expressed in the mass basis couplings in Eq.\ref{leptoquark_massbasis}.

\subsection{$R_{K^{(*)}}$}

To explain $R_{K^{(*)}}$ with a leptoquark $S_3$ we have \cite{Varzielas:2015iva, Hiller:2017bzc, Hiller:2018wbv}:
\be
\label{RKuirement}
\lambda_{b\mu} \lambda_{s\mu}^* - \lambda_{be}\lambda_{se}^* \simeq \frac{1.1 M^2}{(35~\text{TeV})^2} \,,
\ee
where $M$ here is the mass of the leptoquark.
In our model, we use Eq.\ref{leptoquark_model} to recast at leading order this requirement as
\be
\left( \frac{x^{Q}_{3}\langle \phi \rangle }{M^{Q}_{4}} \right)^2
\left(|\lambda^{Q_4}_{L_2}|^2  - |\lambda^{Q_4}_{L_1}|^2 \right) V_{ts} V_{tb}^* 
\simeq \frac{1.1 M^2}{(35~\text{TeV})^2} \,.
\ee
This predicts a CKM suppression through $V_{ts}$. 
If we set all the order unity coefficients to $1$, taking care that in the more rigorous approach discussed in Eq. \ref{rigorous_angle}, $\frac{x^{Q}_{3}\langle \phi \rangle }{M^{Q}_{4}}$ should be taken no larger than $1/\sqrt{2}$
, we would have a natural expectation that, for renormalisable couplings $\lambda^{Q_4}_{L_1}$ and $\lambda^{Q_4}_{L_2}$ both of order unity, the mass of the leptoquark $M \sim 4.7$ TeV:
\be
\label{M_RK}
M \simeq \sqrt{|\lambda^{Q_4}_{L_2}|^2  - |\lambda^{Q_4}_{L_1}|^2} \, 4.7 \, \text{TeV} \,.
\ee

\subsection{$B_s - \bar{B}_s$ mixing}

The mass of the leptoquark $M$ is further constrained from $B_s - \bar{B}_s$ mixing, which depends on the same leptoquark couplings required to explain $R_{K^{(*)}}$. The strongest upper bound can be obtained from the upper bound on the $B_s - \bar{B}_s$ mixing phase, and can be expressed as \cite{Varzielas:2015iva}:
\be
(\lambda_{se}\lambda_{be}^*+\lambda_{s\mu}\lambda_{b\mu}^*+\lambda_{s\tau}\lambda_{b \tau}^*)^2 \lesssim \frac{M^2}{(17.3 \, \text{TeV})^2} \,.
\ee
We note that this has a quartic (rather than quadratic) dependence on the leptoquark couplings that arises from the respective loop diagram involving the leptoquark. Due to the different dependence on the leptoquark couplings, one can combine the requirement of explaining $R_{K^{(*)}}$ with the requirement that the upper bound on $B_s - \bar{B}_s$ mixing is satisfied, and obtain a model independent upper bound on $M \leq 40$ TeV \cite{Varzielas:2015iva, Hiller:2017bzc, Hiller:2018wbv}. If the leptoquark were to be heavier than this, explaining $R_{K^{(*)}}$ could be done only with couplings that are too large, i.e. for that mass they would push the $B_s - \bar{B}_s$ mixing above the experimental upper bound.

Taking e.g. $M \sim 4.7~\text{TeV}$, the bound corresponding to such a mass is
\be
(\lambda_{se}\lambda_{be}^*+\lambda_{s\mu}\lambda_{b\mu}^*+\lambda_{s\tau}\lambda_{b \tau}^*)^2 \lesssim \left(\frac{4.7}{17.3}\right)^2 \simeq 0.074 \,.
\ee
The present model easily accommodates the $B_s - \bar{B}_s$ mixing bound by about one order of magnitude, as the leptoquark couplings are suppressed due to the CKM suppression coming from $V_{ts}^2$:
\be
(\lambda_{se}\lambda_{be}^*+\lambda_{s\mu}\lambda_{b\mu}^*+\lambda_{s\tau}\lambda_{b \tau}^*)^2 \sim
\left( |\lambda^{Q_4}_{L_1}|^2 + |\lambda^{Q_4}_{L_2}|^2 + |\lambda^{Q_4}_{L_3}|^2 \right)^2  |V_{ts} V_{tb}^*|^2 \sim 10^{-3} \,.
\ee

\subsection{$\mu \to e \gamma$}

The leptoquark couplings are however constrained by lepton flavour violating (LFV) bounds \cite{Varzielas:2015iva}. Particularly stringent are $\mu \to e$ conversion processes, with the current bound on $\mathcal{B}(\mu \to e \gamma) = 5.7 \cdot 10^{-13}$ \cite{Adam:2013mnn}, leading to the constraint:
\be
\label{LFV_mue}
| \lambda_{qe} \lambda_{q \mu}^* | \lesssim \frac{M^2}{(34~\text{TeV})^2} \,,
\ee
which we recast in terms of our model as
\be
\left( \frac{x^{Q}_{3}\langle \phi \rangle }{M^{Q}_{4}} \right)^2 | \lambda^{Q_4}_{L_1} \lambda^{Q_4}_{L_2} V_{tb}^2  | \lesssim \frac{M^2}{(34~\text{TeV})^2} ,.
\ee
Where, if we set all the order unity coefficients to $1$, and remembering that in the more rigorous approach discussed in Eq. \ref{rigorous_angle}, $\frac{x^{Q}_{3}\langle \phi \rangle }{M^{Q}_{4}}$ should be taken no larger than $1/\sqrt{2}$:
\be
M \gtrsim \sqrt{|\lambda^{Q_4}_{L_1} \lambda^{Q_4}_{L_2}|} \, 24 \, \text{TeV} \,.
\ee

Comparing Eq.\ref{RKuirement} to Eq.\ref{LFV_mue} indicates it is desirable to have some hierarchy between $\mu$ and $e$ couplings.
Indeed, if the renormalisable couplings $\lambda^{Q_4}_{L_1}$ and $\lambda^{Q_4}_{L_2}$ are both order unity, then $\lambda_{be}$ and $\lambda_{b\mu}$ are of similar size and this constraint pushes the mass of the leptoquark to be bigger than about $24$ TeV. This is approaching the model independent bound applying to the leptoquark $S_3$ of $M \lesssim 40$ TeV \cite{Varzielas:2015iva, Hiller:2017bzc, Hiller:2018wbv}, coming from $B_s - \bar{B}_s$ mixing combined with Eq.\ref{RKuirement}. More importantly for our model, there is an apparent tension with $M \sim 4.7$ TeV appearing in Eq. \ref{M_RK}.

This contrast is indirectly saying something important about the underlying model. Indeed, the strong predictions of our model come from the simple idea that the leptoquark and SM Yukawa couplings are related. It turns out that in this model, the unsuppressed $x_3^Q$ couplings (the same that are responsible for the top quark Yukawa) are also the ones contributing at leading order to several entries of $\lambda_{dl}$. The requirement from explaining $R_{K^{(*)}}$ and the prediction for $\mu \to e \gamma$ are strongly correlated, involving the same $\left( \frac{x^{Q}_{3}\langle \phi \rangle }{M^{Q}_{4}} \right)^2$ factor and the $\lambda^{Q_4}_{L_1}$ and $\lambda^{Q_4}_{L_2}$. The main difference is that in $R_{K^{(*)}}$, there is a CKM suppression from $V_{ts}$ which is absent in $\mu \to e \gamma$. This in turn indicates that the mixed product $|\lambda^{Q_4}_{L_1} \lambda^{Q_4}_{L_2}|$ should be smaller than the largest of the $|\lambda^{Q_4}_{L_1}|^2$ or $|\lambda^{Q_4}_{L_2}|^2$. The most natural way to obtain this is to assume that there is some hierarchy between the two couplings, namely $|\lambda^{Q_4}_{L_1}| \ll  |\lambda^{Q_4}_{L_2}|$. One simple possibility to realise would be to enforce $\lambda^{Q_4}_{L_1}=0$ by means of an additional family symmetry distinguishing the family eigenstate $L_1$ e.g. by an additional $Z_2'$ (the prime distinguishes this from the already present $Z_2$), which is in fact entirely consistent with Eq.\ref{M^lepton_an_basis},
where $L_1$ was already treated differently. Given that the coupling vanishes in the symmetric limit,
$|\lambda^{Q_4}_{L_1}| \ll  |\lambda^{Q_4}_{L_2}|$ is technically natural.\footnote{
An explicit realisation of this would simply assign $L_1$ as odd under $Z_2'$. As the only field odd under this family symmetry, no couplings with $L_1$ are invariant. To make the setup slightly more realistic, one could add a further scalar charged under $Z_2'$, say $\phi'$, whose VEV breaks this family symmetry. Then, couplings involving $L_1$ always appear with a further suppression by $\sim \langle \phi' \rangle/M'$, where $M'$ could be the mass of some additional $Z_2'$-odd messenger fields. The breaking of $Z_2'$ in such a setup thus links the smallness of $m_e/m_\mu$ with the smallness of $|\lambda^{Q_4}_{L_1}|/|\lambda^{Q_4}_{L_2}|$.
}

\subsection{Other LFV}

Given the large entries in each column at the $b$ row, the model predicts correlated large LFV effects in other processes.
The bounds on LFV with the $\tau$ lepton, $\mathcal{B}(\tau \to e \gamma) = 1.2 \cdot 10^{-7}$ \cite{Hayasaka:2007vc} and $\mathcal{B}(\tau \to \mu \gamma) = 4.4 \cdot 10^{-8}$ \cite{Aubert:2009ag} can in principle constrain the respective leptoquark couplings:
\begin{equation}
\label{LFV_taue}
| \lambda_{qe} \lambda_{q \tau}^* | \lesssim \frac{M^2}{(0.6~\text{TeV})^2} \,,
\end{equation}
\begin{equation}
\label{LFV_taumu}
| \lambda_{q\mu} \lambda_{q \tau}^* | \lesssim \frac{M^2}{(0.7~\text{TeV})^2} \,.
\end{equation}
We predict these in terms of our model parameters, where the dominant contribution comes from the coupling of the leptoquark to $b$ (the coupling to $s$ is CKM suppressed):
\be
| \lambda_{b\tau} \lambda_{be}^* | =  \left( \frac{x^{Q}_{3}\langle \phi \rangle }{M^{Q}_{4}} \right)^2 | \lambda^{Q_4}_{L_3} \lambda^{Q_4}_{L_1} V_{tb}^2 | \lesssim \frac{M^2}{(0.6~\text{TeV})^2} \,,
\ee
\be
| \lambda_{b\mu} \lambda_{b \tau}^* | = \left( \frac{x^{Q}_{3}\langle \phi \rangle }{M^{Q}_{4}} \right)^2 | \lambda^{Q_4}_{L_2} \lambda^{Q_4}_{L_3} V_{tb}^2 | \lesssim \frac{M^2}{(0.7~\text{TeV})^2} \,.
\ee
Clearly, the experimental sensitivity isn't competitive with the bound coming from $\mu \to e \gamma$ , unless we take the
 $|\lambda^{Q_4}_{L_1}| < |\lambda^{Q_4}_{L_2}|$ limit discussed above. In such a situation $\mu \to e \gamma$ can be sufficiently suppressed (and so is $\tau \to e \gamma$ in order to leave $\tau \to \mu \gamma$ as the most relevant LFV constraint.

In any case, as the leading contributions for LFV processes depend on the 3 $\lambda^{Q_4}_{L_i}$, the 3 LFV processes are directly related to these 3 renormalisable couplings parameters through their ratios, e.g. the model predicts
\be
\frac{\mathcal{B}(\tau \to e \gamma)}{\mathcal{B}(\mu \to e \gamma)} = \left| \frac{\lambda^{Q_4}_{L_3}}{\lambda^{Q_4}_{L_2}} \right| \,.
\ee

\section{Conclusions \label{sec:conc}}

We have proposed a simple extension of the Standard Model in Table~\ref{tab:funfields1} involving a vector-like fourth family of fermions which are distinguished 
from the SM fermions by a single $Z_2$ symmetry under which the Higgs doublet is odd, preventing Yukawa couplings, together with a $Z_2$-odd scalar leptoquark $S_3$, and a scalar $\phi$ whose VEV breaks the $Z_2$.
The explicit masses of the vector-like fourth family are assumed to be of the same order as the VEV
of the $\phi$ field. In this model, effective Yukawa couplings emerge from the diagrams in Fig.~\ref{Fig1} which are very similar to the diagrams in Fig.~\ref{Fig2} which generate effective
leptoquark couplings, which implies that the Yukawa couplings are closely related to the leptoquark couplings. 

We have seen that the model successfully describes quark mixing angles and also implies that the leading order leptoquark couplings are related to CKM matrix elements (particularly $V_{ts}$ and $V_{tb}$), with several entries being suppressed by smaller CKM elements and by naturally suppressed mass ratios (such as $m_\mu/m_\tau$, $m_\tau/m_t$).
The model naturally accounts for the fermion mass hierarchy with the first family being precisely massless without any modification of the model.
Although a massless first family is an excellent approximation, it does call for some further modification of the model, such as for example adding 
a further vector-like family at higher mass, or possibly extra vector-like Higgs states. However such modifications are not expected to significantly perturb the main features 
of the model presented here. This is illustrated by a more explicit suggestion, where the lightest family of charged leptons is distinguished by a further $Z_2'$ symmetry.

We have shown that the simple model can 
account for $R_{K^{(*)}}$ while easily avoiding the $B_s - \bar{B}_s$ mixing bound. 
However, we have seen that an important constraint from $\mu \to e \gamma$ strongly constrains the product of two renormalisable couplings of the model. 
We have discussed what this important constraint means in the context of the model and discuss how it can be interpreted in terms of a symmetry distinguishing the lightest family of leptons (e.g. the $Z_2'$ mentioned above), again illustrating the possible consequences of the link between Standard Model and leptoquark Yukawa couplings. The further implications for other lepton flavour violating processes have also been discussed. 

\subsection*{Acknowledgements}

IdMV thanks the University of Basel for hospitality.

IdMV acknowledges funding from the Funda\c{c}\~{a}o para a Ci\^{e}ncia e a Tecnologia (FCT) through
the contract IF/00816/2015 and partial support by Funda\c{c}\~ao para a Ci\^encia e a Tecnologia
(FCT, Portugal) through projects CFTP-FCT Unit 777 (UID/FIS/00777/2013), CERN/FIS-PAR/0004/2017 and PTDC/FIS-PAR/29436/2017  which are partially funded through POCTI (FEDER), COMPETE, QREN and EU.

S.\,F.\,K. acknowledges the STFC Consolidated Grant ST/L000296/1 and the European Union's Horizon 2020 Research and Innovation programme under Marie Sk\l{}odowska-Curie grant agreements Elusives ITN No.\ 674896 and InvisiblesPlus RISE No.\ 690575.

\end{document}